\documentclass[a4paper,12pt]{article}

\usepackage{graphics,psfrag}
\usepackage{amsmath,amsthm,amssymb,epsfig,euscript,array,cite,cancel,color}
\usepackage{mathrsfs}
\usepackage{cases,empheq}
\usepackage[colorlinks=false]{hyperref}
\usepackage{verbatim}
\usepackage{ulem}
   \usepackage[vcentermath]{youngtab}   
   
   \usepackage[usenames,dvipsnames]{xcolor}  
\usepackage[font=small,margin=1cm]{caption} 

\usepackage[paper=a4paper,dvips,top=2.54cm,left=2.74cm,right=2.34cm,
    foot=1cm,bottom=2.54cm]{geometry}

\newcommand{\bm}[1]{\mbox{\boldmath$#1$}}
\newcommand{\nn}{\nonumber}

\newcommand{\lrbrk}[1]{\left(#1\right)}
\newcommand{\lrsbrk}[1]{\left[#1\right]}

\newcommand{\vf}{\ensuremath{\bm{\epsilon}}}

\newcommand{\Lf}[1]{\ensuremath{\bm{L}_{#1}}}

\newcommand{\df}{\ensuremath{\bm{\mathrm{d}}}}

\newcommand{\Ef}{\ensuremath{\bm{E}}}\tabularnewline

\def \be {\begin{equation}}
\def \ee {\end{equation}}
\def \bea {\begin{eqnarray}}
\def \eea {\end{eqnarray}}
\newcommand{\eq}[1]{(\ref{#1})}

\author{Baoyi Chen\footnote{baoyi@tapir.caltech.edu}$~^a$, 
Feng-Li Lin\footnote{linfengli@phy.ntnu.edu.tw}$~^b$,
Bo Ning\footnote{ningbo@scu.edu.cn}$~^c$
and  Yanbei Chen\footnote{yanbei@catech.edu}$~^a$
\\
\\
{\small\it $^a$ TAPIR, Walter Burke Institute for Theoretical Physics,} 
\\
{\small\it  California Institute of Technology, Pasadena, California 91125, USA}
\\
{\small\it $^b$ Department of Physics, National Taiwan Normal University,}
\\
{\small\it No. 88, Sec. 4, Ting-Chou Road, Taipei 11677, Taiwan}
\\
{\small\it $^c$  College of Physical Science and Technology, }
\\
{\small\it Sichuan University, Chengdu, Sichuan 610065, China}
}

\title{\bf {\Large Constraints on low-energy effective theories from weak cosmic censorship} }

\date{}
\begin{document}

\maketitle

\thispagestyle{empty}

\abstract{We examine the weak cosmic censorship conjecture (WCCC) for the extremal charged black hole in possible generalizations of Einstein-Maxwell theory due to the higher order corrections, up to fourth-derivative terms.
Our derivation is based on Wald's gedanken experiment to destroy an extremal black hole. 
We find that, provided the null energy condition for the falling matter, the WCCC is preserved for all possible generalizations.  Thus, the WCCC cannot serve as a constraint to the higher order effective theories.  We also show that up to first order variations of black hole mass and charge, WCCC is preserved for nonrotating extremal black holes in all $n$-dimensional diffeomorphism-covariant theories of gravity and $U(1)$ gauge field.
}

\thispagestyle{empty}
\newpage

\section{Introduction}
Even though the curvature singularity of a black hole is hidden behind the horizon,  it might still be possible to throw charged or spinning matter into a black hole in particular ways that  can destroy the horizon, revealing the singularity previously hidden inside.
This kind of gedanken experiments was first proposed long ago by Wald \cite{Wald:I} to test the so-called Weak Cosmic Censorship Conjecture (WCCC) \cite{Penrose:1969pc}, which asserts that the above gedanken experiments cannot succeed in order to prevent the singularity from being visible. Although the WCCC can be checked easily for extremal black holes,  
it is nontrivial to prove for near-extremal black holes \cite{Wald:I,Hubeny:1998ga} and for general forms of matter.  Recently, significant progress for the general proof of the WCCC has been made by Sorce and Wald  \cite{Sorce:2017dst} who adopted a general relativistic formulation of the energy conservation which can work for general forms of matter obeying the Null Energy Condition (NEC). In this way, they were able to avoid solving the complicated dynamical problems of the in-falling matter involving the self-force effect, and succeeded to show that the WCCC holds for the black holes in Einstein-Maxwell theory, up to second order variation of the black hole's mass, charge and angular momentum. Moreover, their method of examining the WCCC also provides a systematic framework for general theories other than Einstein-Maxwell.

One compelling reason to examine the WCCC for more general theories of gravity and electromagnetism is that the standard Einstein-Maxwell theory, which can be a good approximation at low energies, may need to be corrected at higher energies.
In the low-energy Effective Field Theory (EFT), these quantum corrections can leave low-energy relics in the form of higher-order derivative terms beyond Einstein-Maxwell terms, modifying the black hole solutions, as well as the relativistic laws of the energy-momentum conservation. These terms may also make the WCCC fail. 
If we take the WCCC as a universal physical principle, then only those higher order EFTs that admit the WCCC  should be accepted.
This is in a similar spirit of using the weak gravity conjecture \cite{Vafa:2005ui,ArkaniHamed:2006dz,Cheung:2018cwt} which takes  ``gravity force is the weakest in nature'' as a new physical principle to constrain the higher order EFTs \cite{ArkaniHamed:2006dz}.  In this paper, we would like to check to see if weak cosmic censorship can serve as a constraint on the EFTs with higher-derivative terms.  By the end we will show that there is no constraint WCCC can put on the coupling coefficients of such EFTs. This calls for the further examination of preserving WCCC at the second order variation for the EFTs considered here.     
  
 {\bf Note added:} A technical error regarding \eq{eq:gauge-sols} in the previous version of this work is pointed out by \cite{Jiang:2021ohh}. In this version we fix the error and reach the same results as in  \cite{Jiang:2021ohh}.

\section{EFTs, Black-Hole Solutions, and Extremality Condition} 
 To demonstrate the power of the WCCC as a constraint to the EFTs, in this work we consider the most general quartic order corrections to Einstein-Maxwell theory, which is given by the following EFT action:
\begin{equation}
 \label{eq:action}
  I= \int d^4{}x \; \sqrt{-g}{}(\frac{1}{2\kappa}R -\frac{1}{4}F_{\mu\nu}F^{\mu\nu}+\Delta L ) \,,
\end{equation}
where  \footnote{We have neglected terms proportional to $\nabla^\mu F_{\mu \rho}\nabla_\nu F^{\nu \rho}$, as it does not affect the black hole metric or our parameter bound.  Further note that terms like
$(\nabla_{\mu}F_{\nu\rho})(\nabla^{\mu}F^{\nu\rho})$ and
$(\nabla_{\mu}F_{\nu\rho})(\nabla^{\nu}F^{\mu\rho})$ can be recasted (up to some constant factor)
into $\nabla^\mu F_{\mu \rho}\nabla_\nu F^{\nu \rho}$ plus existing terms in $\Delta L$ and an additional boundary term,
upon using Bianchi identities, Ricci identities, and integrating by
parts~\cite{Deser:1974cz}.  
}
\begin{align}
 \Delta  L  =\quad & c_1R^2+ c_2 R_{\mu\nu}R^{\mu\nu}+c_3 R_{\mu\nu\rho\sigma}R^{\mu\nu\rho\sigma}\;  \nn \\ \nn
  +\, &c_4R F_{\mu\nu}F^{\mu\nu}+c_5R_{\mu\nu}F^{\mu\rho}F^{\nu}{}_{\rho} +
        c_6R_{\mu\nu\rho\sigma}F^{\mu\nu}F^{\rho\sigma} \;\\  
  +\, & c_7 F_{\mu\nu}F^{\mu\nu}F_{\rho\sigma}F^{\rho\sigma} + c_8
        F_{\mu\nu}F^{\nu\rho}F_{\rho\sigma}F^{\sigma\mu} 
        \,.
\end{align}
We will assume $c_i$'s are small and restrict our consideration to ${\cal O}(c_i)$. The aforementioned self-interactions of four photons are the terms with coupling coefficient $c_7$ and $c_8$ respectively.

For simplicity, we will consider only the charged non-spinning black holes.  
The perturbative procedure of solving such black hole solutions has been outlined in \cite{Kats:2006xp}, leading to {\it a family} of solutions parametrized by the mass and the charge $(M,Q)$.  Here we list some partial results relevant for our considerations~\footnote{See Sec. I of the supplemental materials for the full explicit expressions.}, namely the Maxwell gauge field
\begin{align}
\label{eq:gauge-sols}
A_t = - \frac{q}{r} - \frac{q^3}{5r^5} \times 
  \lrsbrk{c_2 \kappa ^2 + 4 c_3 \kappa ^2 + 10 c_4 \kappa + c_5 \kappa - c_6 \kappa\lrbrk{9-\frac{10mr}{q^2}} - 16 c_7-8 c_8}\,,
\end{align}
and the $tt$-component of the metric~\footnote{See Sec. II of the supplemental materials for the full expressions.} 
\begin{align}\nn
-g_{tt} =& 1-\frac{\kappa  m}{r}+\frac{\kappa  q^2}{2 r^2} + 
c_2 \left(\frac{\kappa ^3 m q^2}{r^5}-\frac{\kappa ^3 q^4}{5 r^6}-\frac{2 \kappa ^2 q^2}{r^4}\right) \\ \nn
 &+c_3 \left(\frac{4 \kappa ^3 m q^2}{r^5}-\frac{4 \kappa ^3 q^4}{5 r^6}-\frac{8 \kappa ^2 q^2}{r^4}\right) \\ \nn
 &+c_4 \left(-\frac{6 \kappa ^2 m q^2}{r^5}+\frac{4 \kappa ^2 q^4}{r^6}+\frac{4 \kappa  q^2}{r^4}\right) \\ \nn
 &+c_5 \left(\frac{4 \kappa ^2 q^4}{5 r^6}-\frac{\kappa ^2 m q^2}{r^5}\right) \\ \nn
 &+c_6 \left(\frac{\kappa ^2 m q^2}{r^5}-\frac{\kappa ^2 q^4}{5 r^6}-\frac{2 \kappa  q^2}{r^4}\right) \\ 
 &+ c_7 \lrbrk{-\frac{4 \kappa  q^4}{5 r^6}}+ c_8 \lrbrk{-\frac{2 \kappa  q^4}{5 r^6}} +{\cal O}(c_i^2)\,. \label{corrected-TT}
\end{align}
Here we define the reduced mass $m\equiv M/4\pi$, the reduced charge $q\equiv Q/4\pi$ and $\kappa=8\pi G_N$, where $G_N$ is the gravitational constant. Note that in \eq{corrected-TT} there is no ${\cal O}(c_1)$ correction.

As shown by Ref.~\cite{Kats:2006xp}, as long as
\begin{equation}
\label{MQbound}
m\ge \sqrt{\frac{2}{\kappa}} \vert q \vert   \left(1- \frac{4}{5q^2} c_0 \right) \,,
\end{equation}
the singularity of the space-time will be hidden by a horizon; more precisely, the {\it outer horizon} located at the outer most solution of $g_{tt}(r_H)=0$. Here 
\begin{equation} \label{eq:c0}
c_0 \equiv  c_2 + 4c_3 + \frac{c_5}{\kappa} + \frac{c_6}{\kappa} + \frac{4 c_7}{\kappa^2} + \frac{2 c_8}{\kappa^2}\,,
\end{equation}
and $c_0\rightarrow 0 $ recovers the Reissner-Nordstrom solution of Einstein-Maxwell. 
For a fixed $m$, as $q$ increases to, and then exceeds, the critical value at which equality holds in~\eqref{MQbound}, two horizons  will merge and subsequently disappear, revealing the singularity.  In this way, the {\it extremal solution} is defined by imposing equality in \eqref{MQbound}.  This implicitly defines a function $q_{ext}(m)$  for the extremal solution.  For each  $m$, the horizon radius of the extremal solution is given by
\begin{align}\label{newrH}
r_H^{ext}={m \kappa \over 2} +\frac{4}{5m} \lrbrk{c_2 + 4c_3 +\frac{10 c_4+c_5+c_6}{\kappa} -\frac{16c_7+8c_8}{\kappa^2}}.
\end{align}
On this extremal horizon, the electrostatic potential is  
\be\label{Phic}
\Phi_{H}^{ext}= - \lrbrk{\xi^a A_a}\vert_{\mathcal{H}} =  \sqrt{\frac{2}{\kappa}} \lrbrk{1 + \frac{4 c^\prime_0}{5q^2} } \,,
\ee
where $\vec \xi = \vec\partial_t$ is the time-like Killing vector of the space-time, and 
\begin{equation}
c^\prime_0 = c_2 + 4c_3 + \frac{c_5}{\kappa} + \frac{c_6}{\kappa} + \frac{4 c_7}{\kappa^2} + \frac{2 c_8}{\kappa^2} \,.
\end{equation}
One immediately notes that $c^\prime_0 = c_0$, but we shall discuss the conseqeunce later.
We refer to $(m,q)$ solutions that strictly satisfy the inequality \eqref{MQbound} as {\it regular solutions}, those that take equality as {\it extremal solutions}, and those that violate the inequality as  {\it singular solutions}.  We may still refer to them as ``black holes'', even though the horizon may or may not be destroyed.

\section{Gedanken Experiment to Destroy the Horizon}
In gedanken experiments that attempt to destroy the horizon, e.g., as set up by Wald~\cite{Wald:I,Sorce:2017dst}, we shall always (if tacitly) assume stability of our family of solutions.  That is, starting off with a regular solution $(m,q)$, as we ``throw matter into'' it, the final space-time geometry and field configuration will settle down to another solution {\it in our family}.
If our ``way of throwing matter'', for example described by  the on-shell metric perturbations, field perturbations and matter stress-energy tensor in the initial slice,  is parameterized by $w$, then the final solution should be given by $(m(w),q(w))$. 

In this language, the WCCC dictates that a starting regular solution $(m,q)$ long before ``throwing matter'' will only lead to $(m(w),q(w))$ that are still regular.  As a special case, let us now consider a starting extremal solution $(m,q_{ext}(m))$, and a particular approach of throwing matter, we can write 
\begin{equation}
m(w) = m + w \delta m +O(w^2)\,,\quad q(w) = q_{ext}(m) + w\delta q+O(w^2) \,.
\end{equation}
The condition for the starting extremal solution to not become singular, at first order in $w$, is given by 
\begin{equation}
\label{eq:wccc-m-q-relation}
\delta m - \sqrt{\frac{2}{\kappa}} \lrbrk{1 + \frac{4 c_0}{5q^2} }\delta q \geq 0\,.
\end{equation}
We therefore need to find out whether physical laws in our modified theory imposes that \eqref{eq:wccc-m-q-relation} must hold for all infalling matter  --- or to find a particular way of throwing matter that violate \eqref{eq:wccc-m-q-relation}.  The advantage of starting-off at the extremal solution is: once Eq.~\eqref{eq:wccc-m-q-relation} is violated, then any infinitesimal $w$ will lead to destruction of the horizon, and we can restrict ourselves to linear perturbation. 

By contrast, starting from a non-extremal black hole with $(m,q_{ext}(m)-\epsilon)$, a {\it finite step size} for $w$ must be made to surpass the extremality contour, and in this case the higher derivatives of $m(w)$ and $q(w)$ may become important, requiring the computation of higher-order variations. 
 This was indeed the situation encountered by Hubeny \cite{Hubeny:1998ga}, which was latter addressed by Sorce and Wald  \cite{Sorce:2017dst}  by considering the second order variations.  Intuitively, one would expect the sub-extremal black holes will obey the WCCC if the extremal ones do, but the second order variations are needed for a rigorous examination on the sub-extremal case. In this paper we shall restrict ourselves to the extremal black holes.

As it turns out, condition \eqref{eq:wccc-m-q-relation} coincides with the requirement that the horizon area must increase as matter fall into extremal black holes~\footnote{See Sec. IV of the supplemental materials for details}. More specifically, if we denote by $\mathcal{A}(m,q)$ the area of the horizon, then one can show that 
\begin{equation}
\partial_m  \mathcal{A}(m,q) /\partial_q  \mathcal{A}(m,q)|_{q=q_{ext}(m)} =  d q_{ext}(m)/dm \,,
\end{equation}
and that $ d\mathcal{A}(m+w dm,q_{ext}(m)+w dq)/dw = 0$ is equivalent to the equality in Eq.~\eqref{eq:wccc-m-q-relation}. In this way, the violation of condition~\eqref{eq:wccc-m-q-relation}, or the {\it destruction of the extremal horizon}, relies on the possibility of {\it area decrease} at linear order.   This can be possible for the theories we consider even when the NEC is satisfied, because Raychaudhuri equation is now modified, and the NEC does not always lead to attractive gravity.



\section{Test Particle}
For a regular solution $(m,q)$, consider a test particle with  reduced mass $\delta m_0$ and reduced charge $\delta q_0$, falling in from infinity. Using  the {\it minimally coupled} action of  
\begin{equation}
 S_p =   4 \pi \int d\tau \, (\delta m_0  -\delta q_0 \vec u \cdot \vec A ) \,,
\end{equation}
the  reduced canonical momentum of the particle,  $\vec p = \delta m_0 \,\vec u -\delta  q_0\,\vec A$, satisfies $\vec\xi\cdot\vec p =\mathrm{const}$ along the particle's trajectory;  at linear order in $\delta m_0$ and $\delta q_0$, we do not have to consider the radiation reaction. Applying this to the particle at infinity and  on the horizon, we obtain
\begin{equation}
\delta m_0 \Big (\vec u^{\,H} \cdot\vec\xi \,\Big) -\Phi_H^c \delta q_0 = \delta m_0 \Big (\vec u^{\,\infty}\cdot\vec\xi\, \Big) = -\delta  E_\infty \,,
\end{equation}
where $\vec u^{\,\infty}$  and $\vec u^{\,H}$ are the 4-velocities of the particle at infinity and on the horizon, and we have used the fact that $A_t$ does not depend on $t$, hence $\vec \xi\cdot\vec A$ vanishes at infinity.   

For the final space-time, assuming that it still belongs to the same family, with $(m+\delta m,q+\delta q)$.  We can argue from the charge conservation that $\delta q =\delta q_0$, and, from the conservation of ADM mass, as well as the fact that the energy of gravitational radiation emitted by the in-fall process is $\mathcal{O}(\delta m^2)$, that $\delta m = \delta E_\infty$: basically, {\it the charge and the energy of the particle are added to those of the black hole}.  We will soon give a more rigorous justification, but with this in hand we can write 
\begin{equation}
\label{particlerelation}
\delta m  -\Phi_H^c \delta q =- \delta m_0 \Big (\vec u^{\,H} \cdot\vec\xi \,\Big) \ge 0  \,.
\end{equation}
The latter inequality is because $\vec u^{\,H}\cdot\vec \xi\le 0$: the 4-velocity of the particle must be pointed toward the future as the particle crosses the horizon.  This can be saturated if the particle is able to ``rest right on top of the horizon''.   Inserting Eq.~\eqref{Phic} into Eq.~\eqref{particlerelation}, we obtain the relation between $\delta q$ and $\delta m$ in this in-falling test particle situation: 
\begin{equation}
\label{condparticle}
\delta m \ge \sqrt{\frac{2}{\kappa}}\left(1+\frac{4c_0'}{5q^2}\right) \delta q  \,.
\end{equation}
This is clearly the same as Eq.~\eqref{eq:wccc-m-q-relation} since $c^\prime_0 = c_0$.  However, before discussing its consequences, we shall introduce the framework by  Sorce and Wald, which provides more rigorous treatment of the energy conservation, and is able to treat more general infalling matter.

\section{Sorce-Wald method for generic matter}
\begin{figure} 
  \centering
  \includegraphics[width=0.75\linewidth]{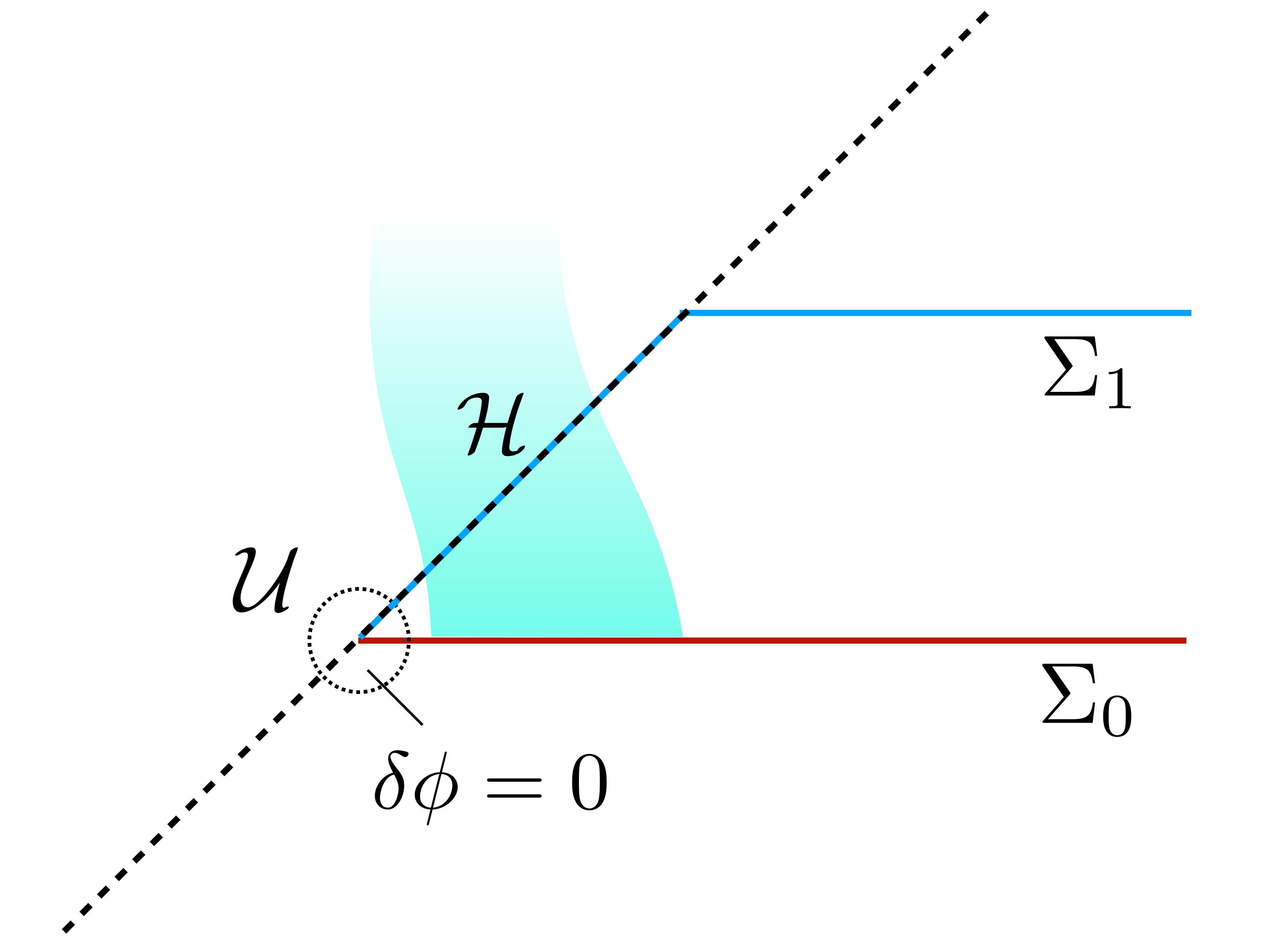}
  \caption{The gedanken experiment to destroy an extremal black hole.  Charged matter, occupying the shaded region, crosses the $\mathcal{H}$ portion of the extremal horizon.}
  \label{fig:gedanken-plot}
  \end{figure}

%
We now sketch the method of Sorce and Wald developed in \cite{Sorce:2017dst,Hollands:2012sf}. 
We follow the notation of Wald, and denote by $\phi =(g_{ab},A_a)$ the metric and field degrees of freedom.  We start off with an extremal black hole, with $(m,q_{ext}(m))$, and define a Cauchy surface $\Sigma_0$ at early time, and a hypersurface $\Sigma_1$ which starts at sufficiently late time when the matter all fall in, and terminates at null infinity. We denote by $\mathcal{H}$ the portion of the extremal horizon between $\Sigma_0$ and $\Sigma_1$ (see Fig.~\ref{fig:gedanken-plot}). We then apply perturbation $\delta \phi$, as well as matter, with  stress-energy tensor $\delta T_{ab}$ and electric current $\delta j_a$, also a form of perturbation, in an open neghborhood of $\Sigma_0$.  We will setup our initial value problem in such a way that $\delta\phi$, $\delta T_{ab}$ and $\delta j_a$ all vanish in an open neighborhood $\mathcal{U}$ surrounding the intersection of $\mathcal{H}$ and $\Sigma_0$.  In principle, $\delta \phi$ and $\delta T_{ab}, \,\delta j_a$ should be evolved jointly into the future, but here we assume stability of our family of solutions, and therefore can impose that space-time geometry  in an open neighborhood of  $\Sigma_1$ is that of $(m+\delta m,q+\delta q)$ \footnote{We note that  in general $\Sigma_1$ is only a portion of a ``Cauchy surface" --- with the remaining portion completed by a portion of the future null infinity.   A Cauchy surface that ends at spatial infinity does not approach $(m+\delta m,q+\delta q)$, fast enough; its ADM mass is not equal to $m+\delta m$ either, because it contains the energy-momentum content of gravitational waves emitted during the in-fall process, see e.g. \cite{Hollands:2012sf}. Sorce and Wald simply assumed that the late time solution is stable and non-radioactive to bypass the above concern \cite{Sorce:2017dst}. On the other hand, in this paper we consider only the infall of matter into an extremal black hole for which the dynamics is non-radioactive, thus the above issue does not exist in our consideration.}. 

  A general Noether method to derive the law of energy conservation for such an in-falling process is developed by Iyer and Wald \cite{Iyer:1994ys}, which we will briefly sketch as follows. Given a theory Lagrangian $L(\phi)$ of gravity and matter, we can introduce the Lagrangian 4-form $\Lf{} =L \vf$, 
  where $\vf$ is the volume form associated with the metric. Then,  variation of $\Lf{}$ yields
\be
\delta \Lf{} = {\bf E}(\phi) \delta \phi+ d {\bf \Theta} (\phi,\delta \phi) \,,
\ee   
 where ${\bf E}(\phi)=0$ is Euler-Lagrangian equation, and ${\bf \Theta} (\phi,\delta \phi)$ is the symplectic potential 3-form.   For an arbitrary vector $\xi^a$, one can construct the associated Noether current ${\bf J}_{\xi}={\bf \Theta} (\phi,{\cal L}_{\xi} \phi) - i_{\xi} \Lf{}$, which, because ${\bf J}_{\xi}$ is conserved, i.e., $d{\bf J}_{\xi}=0$,  can be rewritten as ${\bf J}_{\xi}=d {\bf Q}_{\xi}+ \xi^a {\bf C}_a$ with the 3-form constraint ${\bf C}_a=0$ when equations of motion are satisfied. 
For instance, in Einstein-Maxwell theory, the 3-form constraint is given by
\begin{equation}
\label{eq:constraint-EM}
({\bf C}_a)_{bcd} = \epsilon_{ebcd}({T^{e}}_a +j^{\,e} A_a),
\end{equation}
with $T_{ab} = \frac{1}{\kappa} \lrbrk{G_{ab}-\kappa T^{\rm EM}_{ab}}$  the non-electromagnetic stress energy tensor, and $j^{\,a}= \nabla_b F^{ab}$ the charge current of the Maxwell source.   Thus the on-shell condition ${\bf C}_a=0$ gives the equations of motion $G_{ab}=\kappa T^{\rm EM}_{ab}$ and $\nabla_b F^{ab}=0$. The form \eqref{eq:constraint-EM} also holds when the higher-order derivative corrections $\Delta L$ are present.
Assuming ${\bf E}(\phi)=0$ and $\xi^a$ is a Killing vector, i.e., ${\cal L}_{\xi} \phi=0$, it is easy to show that $\delta {\bf J}_{\xi}=d i_{\xi} {\bf \Theta} (\phi,\delta \phi)$ which is then combined with $\delta {\bf J}_{\xi}=d \delta {\bf Q}_{\xi}+ \xi^a \delta {\bf C}_a$, and is integrated over the hypersurface ${\cal H} \cup \Sigma_1$ to yield
\be\label{1st-variation-proto}
\int_{\infty} [ \delta {\bf Q}_{\xi}-i_{\xi} {\bf \Theta} (\phi,\delta \phi) ]=-\int_{{\cal H} \cup \Sigma_1} \xi^a \delta {\bf C}_a \,,
\ee 
where we have used the Stoke's theorem to turn the 3-surface integral into the boundary integrals at the spatial infinity  $\infty$ and at the intersection $\mathcal{H} \cap \Sigma_0$, by also imposing $\delta \phi=0$ at $\mathcal{H} \cap \Sigma_0$. 

If we assume $\xi^a$ is the time-like Killing vector $t^a =(\partial_t)^a$ for non-spinning black holes, then we denote the change of the ADM mass as
\be\label{deltaM-1}
\delta {\cal M}=\int_{\infty} [ \delta {\bf Q}_{\xi}-i_{\xi} {\bf \Theta} (\phi,\delta \phi) ] \,,
\ee
and the charge crossing the horizon as,
\be\label{deltaQ-1}
\delta {\cal Q}\equiv \int_{\mathcal{H}}\vf_{abcd} \, \delta j^{\,a}\,,
\ee
where the electric current $\delta j^{\,a}$ and the stress tensor $\delta{T^a}_b$ can be read off from the following on-shell relation \footnote{To arrive \eq{on-shell-Cab} we have imposed the on-shell conditions for the theory \eq{eq:action} along with the additional minimally coupled matter of stress tensor $\delta T^{ab}$ and charge current $\delta j^a$.} 
\be\label{on-shell-Cab}
(\delta {\bf C}_a)_{bcd}= \vf_{ebcd}  \lrbrk{\delta{T^e}_a + A_a \delta j^e }\,.
 \ee

Combine all above and require vanishing of $\delta j^e$ and $\delta{T^e}_a$ on $ \Sigma_1$ as depicted in Fig. \ref{fig:gedanken-plot}, we can turn \eq{1st-variation-proto} into the the following law of energy conservation for the in-falling process of Wald's gedanken experiment, 
\begin{equation}\label{1st-law}
 \delta {\cal M} - \Phi_{H}^c \delta {\cal Q}  \, = \,-\int_{\mathcal{H}} \vf_{ebcd} \, \xi^a \delta{T^e}_a \,.
\end{equation}
On horizon $\cal H$ we can relate the 4-volume form $\vf$ to the 3-volume form $\tilde{\vf}$  by the relation  $\vf_{ebcd} = - 4 n_{[e}\tilde{\vf}_{bcd]}$ where $n^e$ is the null vector normal to $\cal H$.  Using this relation and the fact $\xi^a \propto n^a$ on $\cal H$,  the R.H.S. of \eq{1st-law} turns into $4 \int_{\cal H} \tilde{\vf} \delta T_{ab} n^a n^b$, which is non-negative if matter's stress tensor obeys the NEC. Thus, the variational identity \eq{1st-law} becomes an inequality for matter obeying the NEC,
\be\label{1st-ineq}
\delta {\cal M}-\Phi_{H}^c \delta {\cal Q}\ge 0 \,.
\ee 
This inequality serves as a constraint on the changes of the black hole's mass and charge for the in-falling process, and will be used to check the WCCC by comparing with the condition \eq{eq:wccc-m-q-relation}.

        

\section{Parameter bounds from WCCC}
The Noether method by Iyer and Wald provides a systematic way to calculate $\delta {\cal M}$ of \eq{deltaM-1} and $\delta {\cal Q}$ of \eq{deltaQ-1} for general theory by evaluating $\bf \Theta$, $\bf Q$ and ${\bf C}_a$. For example, these quantities for Einstein-Maxwell theory have been derived in \cite{Iyer:1994ys}, and the results $\delta {\cal M}=4\pi \delta m$ and $\delta {\cal Q}=4\pi \delta q$ are then used to show that the WCCC holds for Einstein-Maxwell theory.


 Here we apply the same method for our higher-order theory   \eq{eq:action}. The derivation is tedious but straightforward, and the result is given in the supplemental materials~\footnote{See Sec. III of the supplemental materials for the explicit expressions.}, based on which we can evaluate the corresponding $\delta {\cal M}$ and  $\delta {\cal Q}$.  As a result, we find that $\delta {\cal M} = 4\pi \delta m$ because the corrections due to higher order Lagrangian $\Delta L$ fall off too quickly  to contribute asymptotically to $\delta {\cal M}$. Similarly, we arrive $\delta {\cal Q}=4 \pi \delta q+{\cal O}(c_i^2)$  after tedious calculations~\footnote{This can also be seen as follows. By the construction of source theory, $j_a=\nabla^b(F_{ab}-S_{ab})$ in which  $S_{ab}$ is given in (5) of supplemental materials and is of ${\cal O}(c_i)$, and using (3) of supplemental materials,  $F_{ab}=F_{ab}^{(0,j)}+S_{ab}+{\cal O}(c_i^2)$ where the superscript $(0,j)$ means to evaluate by plugging the background Reissner-Nordstr\"{o}m configurations and keeping up to ${\cal O}(c_i)$ terms. We then arrive $j_a= \nabla^bF_{ab}^{(0,j)}+{\cal O}(c_i^2)$, and use the Gauss's law the integral ${\cal Q}=\int_{\mathcal{H}}\vf_{abcd} \,  j^a =\int_B *F^{(0,j)}+{\cal O}(c_i^2)$. Then, $\delta {\cal Q}=\int_B (*F^{(0,j)}-*F^{(0)})+{\cal O}(c_i^2)=\delta Q+{\cal O}(c_i^2)$, where $\delta Q=4\pi \delta q$ is the charge carried by the in-falling matter. Thus, $\delta {\cal Q}=4 \pi \delta q+{\cal O}(c_i^2)$ is obtained. }. The results are consistent with the test particle case.  Therefore, we conclude that \eq{1st-ineq}, which holds for general forms of matter obeying the NEC, gives the same condition Eq.~\eqref{condparticle} as for the test particle. 

   Compare the energy condition \eq{condparticle} and the WCCC condition \eq{eq:wccc-m-q-relation}, it is not hard to see that we must have $c^\prime_0 \geq c_0$ for the WCCC to hold for theory \eq{eq:action}. 
In our case, we simply have $c^\prime_0 =  c_0$, thus the WCCC is preserved no matter how we choose the coupling coefficients $c_i$.  Thus there is no bound that one can put on these coefficients using the WCCC.

\section{Extension to more general theories}  
 
We have shown that for the effective field theory with higher derivative terms, the WCCC is always preserved and it yields no bound on the coupling coefficients of the theory.  This conclusion is still interesting in the following two aspects: (i)  for near-extremal black holes, our result calls for check of WCCC at the quartic order for low energy effective theories (EFTs) considered in our paper; (ii) for extremal black holes, one might wonder whether WCCC holds for all  other EFTs in general, i.e., whether $c_0'$ coincides with $c_0$ for a reason.  

While  more technical work is needed to clarify (i),  definite conclusions can be drawn for (ii).  More specifically,  
one can prove that WCCC is preserved for nonrotating extremal black holes in all $n$-dimensional diffeomorphism-covariant theories of gravity and $U(1)$ gauge field; this has also been done by Ref.~\cite{Jiang:2021ohh}.

As has been correctly argued in the paper,  in order for matter to fall into the black hole we must have
\begin{equation}\label{infall}
 \delta M - \Phi_H \delta Q \geq 0
\end{equation}
assuming that  the infalling matter obeys null energy condition. On the other hand,  the condition for the extremal solution to not become singular, i.e, to preserve the WCCC , is given by
\begin{equation}\label{WCCC}
\delta M - \lrbrk{\frac{d M}{d Q}}_{\text{ext}} \delta Q \geq 0 \,,
\end{equation}
where $\lrbrk{{d M}/{d Q}}_{\text{ext}}$ is derivative taken along the extremal trajectory in the $(M,Q)$ space. 
Note that this relation is the generic form of Eq.~(14) in our paper.  Here we have also assumed that $(dM/dQ)_{\rm ext} >0$ and that non-extremal black holes have $M>M_{\rm ext}(Q)$. Some example violating these assumptions is discussed in \cite{Jiang:2020alh} , however, the associated physical implication is obscure.

We then write the first law of black hole thermodynamics for non-extremal black holes,
\begin{equation}
\delta M  = T\delta S + \Phi_H \delta Q \,,
\end{equation}
where $S$ in the entropy of the black hole.  As we approach the extremal solution, we have $T\rightarrow 0$, and
\begin{equation}\label{first-law}
\lrbrk{\frac{d M}{d Q}}_{\text{ext}} = \Phi_H\,.  
\end{equation}
This relation can be verified explicitly for the quartic EFTs studied in this paper. Because of \eq{infall} and \eq{first-law}, \eq{WCCC} always holds and concludes our proof. Obviously, the above proof will not work for near-extremal cases because \eq{first-law} does not hold. 

  In conclusion, no matter how we change the action of the EFTs, the condition for the matter falling into the extremal black hole always coincides with the condition for WCCC. Thus, WCCC for extremal black holes will not constrain the form of  low energy EFTs.  For near-extremal black holes, however, one needs to consider second order variations of black hole mass and charge.  It is still possible that after the second order results are taken into consideration, the WCCC only permits a certain region of the parameter space.  We shall leave the second order calculations for the future work.

\subsection*{Acknowledgements}
\noindent {\it Acknowledgements.--} We thank Clifford Cheung for helpful discussions and comments. BC and YC acknowledges the support from the Brinson Foundation, the Simons Foundation (Award Number 568762), and the National Science Foundation, Grants PHY-1708212 and PHY-1708213. FLL is supported by Taiwan Ministry of Science and Technology through Grant No.~106-2112-M-003-004-MY3, and he also thanks the support from NCTS. BN is  supported by the National Natural Science Foundation of China with Grant No.~11975158.

\appendix
\section{Corrections to the Maxwell source and stress tensor}\label{App-A}

We consider the most general fourth-derivative higher order corrections to Einstein-Maxwell theory, namely, \begin{equation}
 \label{eq:action}
  I= \int d^4{}x \; \sqrt{-g}{}(\frac{1}{2\kappa}R -\frac{1}{4}F_{\mu\nu}F^{\mu\nu}+\Delta L)
\end{equation}
where 
\begin{align}
  \Delta L =\quad & c_1R^2+ c_2 R_{\mu\nu}R^{\mu\nu}+c_3 R_{\mu\nu\rho\sigma}R^{\mu\nu\rho\sigma}\; \\ \nn
  +\, &c_4R F_{\mu\nu}F^{\mu\nu}+c_5R_{\mu\nu}F^{\mu\rho}F^{\nu}{}_{\rho} +
        c_6R_{\mu\nu\rho\sigma}F^{\mu\nu}F^{\rho\sigma} \;\\ \nn
  +\, & c_7 F_{\mu\nu}F^{\mu\nu}F_{\rho\sigma}F^{\rho\sigma} + c_8
        F_{\mu\nu}F^{\nu\rho}F_{\rho\sigma}F^{\sigma\mu} 
        \,.
\end{align}

The field equations obtained by the variation of the action \eq{eq:action} with respect to $A_{\mu}$ and $g^{\mu\nu}$ are given respectively by
\begin{equation}
\label{eq:maxwell}
  \nabla_\nu (F^{\mu\nu}-S^{\mu\nu})= 0  \,,
\end{equation}
and   
\begin{equation}
 R_{\mu\nu}-\frac{1}{2}g_{\mu\nu}R =\kappa T_{\mu\nu}
 =\kappa (\tilde{T}_{\mu\nu}+\Delta T_{\mu\nu})\,.
 \label{eq:EFE}
\end{equation}
In the above $\tilde{T}_{\mu\nu}=F_{\mu}{}^\rho F_{\nu\rho} -\frac{1}{4}g_{\mu\nu}F_{\rho\sigma}F^{\rho\sigma}$ is the stress tensor of the Maxwell theory, and $\Delta T_{\mu\nu}$ and $S^{\mu\nu}$ are the corrections respectively to the stress tensor and Maxwell source field from the higher-dimension operators.

Here we list the details of the corrections to the Maxwell source field and stress tensor and , i.e., $S^{\mu\nu}$ in Eq.~(9) and $\Delta T_{\mu\nu}$ in Eq.~(10) of the main text:
\begin{align}\nn
  S^{\mu\nu} = \quad & 4 c_4 RF^{\mu\nu}
                       +2c_5(R^{\mu\rho}F_\rho{}^\nu-R^{\nu\rho}F_\rho{}^\mu)
                       +4c_6R^{\mu\nu\rho\sigma}F_{\rho\sigma} \;+\\
  +\;& 8c_7 F_{\rho\sigma}F^{\rho\sigma}F^{\mu\nu}
       +8 c_8F_{\rho\sigma}F^{\rho\nu}F^{\mu\sigma} \,,
\end{align}
and  
\begin{eqnarray}
  \Delta T_{\mu\nu}\nn
  &=& c_1 \left(g_{\mu\nu} R^2 - 4RR_{\mu\nu} +
      4\nabla_\nu\nabla_\mu R - 4g_{\mu\nu}\Box R \right) + \\ \nn
  &+& c_2 \left(g_{\mu\nu} R_{\rho\sigma} R^{\rho\sigma} +
      4\nabla_\alpha\nabla_\nu R^\alpha_\mu - 2\Box R_{\mu\nu} -
      g_{\mu\nu} \Box R - 4R^\alpha_\mu R_{\alpha\nu} \right) + \\
  \nn
  &+& c_3 \left(g_{\mu\nu} R_{\alpha\beta\gamma\delta}R^{\alpha\beta\gamma\delta} -
      4R_{\mu\alpha\beta\gamma}{R_\nu}^{\alpha\beta\gamma} - 8\Box
      R_{\mu\nu} \right. \\ \nn
      && \left. \quad + 4\nabla_\nu\nabla_\mu R + 8R_\mu^\alpha R_{\alpha\nu}
      - 8R^{\alpha\beta}R_{\mu\alpha\nu\beta} \right) + \\ \nn
  &+& c_4 \left(g_{\mu\nu}R F^2 - 4R{F_\mu}^\sigma F_{\nu\sigma} -
      2F^2R_{\mu\nu} + 2\nabla_\mu\nabla_\nu F^2 - 2g_{\mu\nu}\Box
      F^2\right) + \\ \nn
  &+& c_5 \left(g_{\mu\nu}R^{\kappa\lambda}F_{\kappa\rho}{F_\lambda}^\rho - 4R_{\nu\sigma}F_{\mu\rho}F^{\sigma\rho} -
      2R^{\alpha\beta}F_{\alpha\mu}F_{\beta\nu} )
      \right. \\ \nn 
      && \left. \quad  -
      g_{\mu\nu}\nabla_\alpha\nabla_\beta({F^\alpha}_\rho F^{\beta\rho}+
                          2\nabla_\alpha\nabla_\nu(F_{\mu\beta}F^{\alpha\beta}) -
                          \Box(F_{\mu\rho}{F_\nu}^\rho) \right) + \\ \nn
  &+& c_6 \left(g_{\mu\nu}R^{\kappa\lambda\rho\sigma}F_{\kappa\lambda}F_{\rho\sigma}
      - 6 F_{\alpha\nu}F^{\beta\gamma}{R^\alpha}_{\mu\beta\gamma} -
      4\nabla_\beta\nabla_\alpha({F^\alpha}_\mu{F^\beta}_\nu) \right) +
  \\ \nn
  &+& c_7 \left(g_{\mu\nu}(F^2)^2 - 8F^2{F_{\mu}}^\sigma F_{\nu\sigma}\right) + \\ 
  &+& c_8
      \left(g_{\mu\nu}F^{\rho\kappa}F_{\rho\sigma}F^{\sigma\lambda}F_{\kappa\lambda}
      -8F_{\mu}{}^\rho F_{\nu}{}^\sigma F_{\rho}{}^\kappa
      F_{\sigma\kappa}\right) \,. 
     \label{DeltaT}
     \end{eqnarray}
Note that $F^2=F_{\rho\sigma}F^{\rho\sigma}$ and $\Box = \nabla_a \nabla^a$.

\section{Corrections to the Reissner-Nordstr\"{o}m black hole}\label{App-B}
The functions $\lambda(r)$ and $\nu(r)$ are related to the components
of Ricci curvature tensor $R_{\mu\nu}$ via
\begin{align}
  \frac{1}{2}\left(R^{t}_t-R^{r}_{r}\right) -R^{\theta}_\theta =
  \frac{1}{r^2}\frac{d}{dr}\left[r(e^{-\lambda(r)}-1)\right] \,, \\ \nn
  R^{t}_t-R^{r}_{r}
  = -\frac{e^{-\lambda(r)}}{r}\left[\nu^\prime(r)+\lambda^\prime(r) \right]\,.
\end{align}
To solve for $\lambda$ and $\nu$ explicitly, we need an additional
boundary condition.  Assuming that at $r\rightarrow \infty$ the metric
approaches the Schwarzschld solution, the results are then given by
\begin{align}
  \label{eq:integral-expr}
  e^{-\lambda(r)} &=  1-\frac{\kappa M}{4\pi r} -\frac{1}{r}
                    \int^{\infty}_{r}dr \; r^2\left[ \frac{1}{2}\left(R^{t}_t-R^{r}_{r}\right)
                    -R^{\theta}_\theta \right] \,, \\ \nn
  \nu (r) &= -\lambda(r) + \int^{\infty}_{r} dr \; r
            \left(R^{t}_t-R^{r}_{r}\right)e^{\lambda(r)}\,.
\end{align}
We further take the trace-reverse of Eq.~(10) from the main text and obtain
that
\begin{equation}
  R_{\mu\nu} = \kappa\left(T_{\mu\nu}- \frac{1}{2}T g_{\mu\nu} \right) \,,
\end{equation}
where $T$ is the trace of the total energy-momentum tensor $T_{\mu\nu}$,
and is given by $T = T^{t}_{t}+T^{r}_{r}+2T^{\theta}_{\theta}$.
Plugging the trace-reversed Einstein field equation into the integral
expression~\eqref{eq:integral-expr}, we get
\begin{align}
  e^{-\lambda(r)} &=  1-\frac{\kappa M}{4\pi r} -\frac{\kappa}{r}
                    \int^{\infty}_{r}dr \; r^2 T^t_t \,, \\ \nn
  \nu (r) &= -\lambda(r) +\kappa \int^{\infty}_{r} dr \; r
            \left(T^{t}_t-T^{r}_{r}\right)e^{\lambda(r)}\,.
\end{align}
Once we know the diagonal components of the energy-momentum tensor, it
will be straightforward to compute the corrections to the spherically
symmetric static spacetime as induced by $T_{\mu\nu}$.

We now take our background spacetime to be Reissner-Nordstr\"{o}m
black hole in four-dimension. That is,
\begin{align}
  e^{\nu^{(0)}} = e^{-\lambda^{(0)}} = 1-\frac{\kappa
  M}{4\pi r} + \frac{\kappa Q^2}{32 \pi^2 r^2} \,, \\ \nn
  F^{(0)}_{\mu\nu} dx^{\mu} \wedge dx^{\nu} = \frac{Q}{4\pi r^2}
  dt\wedge dr \,.
\end{align}
Here $\nu^{(0)}(r)$ and $\lambda^{(0)}(r)$ refer to the metric
components in the unperturbed black hole spacetime, and
$ F^{(0)}_{\mu\nu} $ is the background electromagnetic energy-momentum
tensor.  Considering the action in Eq.~(2) of the main text, we treat the
corrections from higher-dimension operators as perturbations.  For
convenience, we also introduce a power counting parameter
$\varepsilon$, and consider a one-parameter family of actions
$I_\varepsilon$, which is given by
\begin{equation}
  I_{\varepsilon}= \int d^4{}x \; \sqrt{-g}{}(L_{0}+\varepsilon\Delta L)\,.
\end{equation}
The original action will be recovered after setting $\varepsilon=1$.
We then expand everthing into powers series in $\varepsilon$.  For
instance,
\begin{align}
  g_{\mu\nu} &= g^{(0)}_{\mu\nu}+\varepsilon h^{(1)}_{\mu\nu} +\mathcal{O}(\varepsilon^2) \,, &
 F_{\mu\nu} &= F^{(0)}_{\mu\nu}+\varepsilon f^{(1)}_{\mu\nu} +\mathcal{O}(\varepsilon^2)\,. &
\end{align}
At order $\varepsilon^1$, the stress energy tensor is given by
\begin{equation}
  T^{(1)}_{\mu\nu} = \widetilde{T}_{\mu\nu}[g^{(0)},f^{(1)},
  F^{(0)}]+  \widetilde{T}_{\mu\nu}[h^{(1)},F^{(0)},
  F^{(0)} ] + \Delta T_{\mu\nu}[g^{(0)},F^{(0)}]\,.
\end{equation}
Noting that in order to compute the corrections to the metric, we
need to calculate $T_{\mu}{}^\nu$ instead of $T_{\mu\nu}$.  At order
$\varepsilon^1$, ${T_{\mu}{}^{\nu}}^{(1)}$ is given by
\begin{equation}
  {T_{\mu}{}^{\nu}}^{(1)} = \widetilde{T}_\mu{}^\nu[g^{(0)},F^{(1)}]+\Delta T_\mu{}^\nu[g^{(0)},F^{(0)}]\,.
\end{equation}
We solve for the corrections to Maxwell equations, and obtain that the gauge field $A_a$ is given by
\begin{align}
A_t = - \frac{q}{r} - \frac{q^3}{5r^5} \times& 
  \lrsbrk{c_2 \kappa ^2 + 4 c_3 \kappa ^2 + 10 c_4 \kappa + c_5 \kappa - c_6 \kappa\lrbrk{9-\frac{10mr}{q^2}} - 16 c_7-8 c_8}\, ,\\
& A_r=A_\theta=A_\phi=0 &\,. 
\end{align}
With the new $A_{\mu}$, we can solve for the corrected energy-momentum tensor ${T_{\mu}{}^{\nu}}^{(1)}$.  We then find the corrected metric tensor component to be
\begin{align}\nn
e^{-\lambda} =& 1-\frac{\kappa  m}{r}+\frac{\kappa  q^2}{2 r^2} + c_2 \left(\frac{3 \kappa ^3 m q^2}{r^5}-\frac{6 \kappa ^3 q^4}{5 r^6}-\frac{4 \kappa ^2 q^2}{r^4}\right) \\ \nn
& + c_3 \left(\frac{12 \kappa ^3 m q^2}{r^5}-\frac{24 \kappa ^3 q^4}{5 r^6}-\frac{16 \kappa ^2 q^2}{r^4}\right)+c_4 \left(\frac{14 \kappa ^2 m q^2}{r^5}-\frac{6 \kappa ^2 q^4}{r^6}-\frac{16 \kappa  q^2}{r^4}\right) \\ \nn
& +c_5 \left(\frac{5 \kappa ^2 m q^2}{r^5}-\frac{11 \kappa ^2 q^4}{5 r^6}-\frac{6 \kappa  q^2}{r^4}\right)+c_6 \left(\frac{7 \kappa ^2 m q^2}{r^5}-\frac{16 \kappa ^2 q^4}{5 r^6}-\frac{8 \kappa  q^2}{r^4}\right)  \\ \nn
&+c_7 \left(-\frac{4  \kappa  q^4}{5 r^6}\right)+c_8\left(-\frac{2  \kappa  q^4}{5 r^6}\right) \,,\nn \\ 
e^{+\nu} =& 1-\frac{\kappa  m}{r}+\frac{\kappa  q^2}{2 r^2} + 
c_2 \left(\frac{\kappa ^3 m q^2}{r^5}-\frac{\kappa ^3 q^4}{5 r^6}-\frac{2 \kappa ^2 q^2}{r^4}\right) \\ \nn
 &+c_3 \left(\frac{4 \kappa ^3 m q^2}{r^5}-\frac{4 \kappa ^3 q^4}{5 r^6}-\frac{8 \kappa ^2 q^2}{r^4}\right)+c_4 \left(-\frac{6 \kappa ^2 m q^2}{r^5}+\frac{4 \kappa ^2 q^4}{r^6}+\frac{4 \kappa  q^2}{r^4}\right) \\ \nn
 &+c_5 \left(\frac{4 \kappa ^2 q^4}{5 r^6}-\frac{\kappa ^2 m q^2}{r^5}\right)+c_6 \left(\frac{\kappa ^2 m q^2}{r^5}-\frac{\kappa ^2 q^4}{5 r^6}-\frac{2 \kappa  q^2}{r^4}\right) \\ 
 &+ c_7 \lrbrk{-\frac{4 \kappa  q^4}{5 r^6}}+ c_8 \lrbrk{-\frac{2 \kappa  q^4}{5 r^6}} \,. \label{DeltaMetric}
\end{align}
In the above we have defined the reduced quantities $m=M/4\pi$ and $q=Q/4\pi$.
Note that the $R^2$-term in the action has no contributions to the equation of motion at leading order in $\varepsilon$. The contributions from $R_{\mu\nu}R^{\mu\nu}$ and $R_{\mu\nu\rho\theta}R^{\mu\nu\rho\theta}$ can be canceled out by choosing $c_2 = -4 c_3$.  This directly confirms that the Gauss-Bonnet term is a topological invariant and does not influence the equation of motion.  Due to the fact that only the $tr$- and $rt-$component of $F_{\mu\nu}$ are nonzero, the term $F_{\mu\nu}F^{\mu\nu}F_{\rho\sigma}F^{\rho\sigma}$ always have twice the contributions from $F_{\mu\nu}F^{\nu\rho}F_{\rho\sigma}F^{\sigma\mu}$ towards the equation of motion.

\section{Explicit forms of ${\bf Q}_{\xi}$ and ${\bf C}_a$ for the higher theory}\label{App-C}

The Lagrangian 4-form $\Lf{}$ for the higher theory can be written as $\Lf{}= \Lf{0} + \sum_i c_i \Lf{i}$.  In this appendix, by following the canonical method developed by Iyer and Wald, we derive and present the Noether charge and constraint associated with each term in $\Lf{}$.

Variation of the Lagrangian 4-form $\Lf{0}$ yields
\begin{equation}
\delta \Lf{0}  = \delta g_{ab}  \lrbrk{-\frac{1}{2\kappa} G_{ab} + \frac{1}{2} T^\mathrm{EM}_{ab}} \vf 
+ \delta A_a \lrbrk{\nabla_b F^{ba}} \vf 
+ \df {\bf \Theta}_0 \,,
\end{equation}
where $G_{ab}=R_{ab} - \frac{1}{2}g_{ab}R$ is the Einstein tensor, and $T^\mathrm{EM}_{ab}$ is the electro-magnetic stress-energy tensor, which is defined by
\begin{equation}
T^\mathrm{EM}_{ab} = F_{ac} {F_b}^c - \frac{1}{4} g_{ab} F_{de} F^{de} \,.
\end{equation}
The symplectic potential can be written as
\begin{equation}
{\bf \Theta}_0 = {\bf \Theta}^{\mathrm{GR}}+ {\bf \Theta}^{\mathrm{EM}} \,,
\end{equation}
where
\begin{align}
\Theta^{\mathrm{GR}}_{abc}\lrbrk{\phi, \delta\phi} & = \frac{1}{2\kappa} \epsilon_{dabc} g^{de}g^{fg} \lrbrk{\nabla_g \delta g_{ef} - \nabla_e \delta g_{fg}}\,, \\
\Theta^{\mathrm{EM}}_{abc} \lrbrk{\phi, \delta\phi} &= - \epsilon_{dabc} F^{de} \delta A_e \,.
\end{align}
Let $\xi^a$ be any smooth vector field on the spacetime.  We find that the Noether charges associated with the vector field are respectively,
\begin{align}
\lrbrk{Q^{\mathrm{GR}}_\xi}_{ab}  & = -\frac{1}{2\kappa} \epsilon_{abcd} \nabla^c \xi^d \,, \\ 
\lrbrk{Q^{\mathrm{EM}}_\xi}_{ab}  & = - \frac{1}{2} \epsilon_{abcd} F^{cd} A_e \xi^e \,.
\end{align}
The equations of motion and constraints are given by
\begin{align}
\Ef_0 \delta \phi &= -\vf \lrbrk{\frac{1}{2} T^{ab} \delta g_{ab} + j^a \delta A_a} \,, \\
C_{bcda} & = \epsilon_{ebcd} \lrbrk{{T^e}_a + j^e A_a} \,,
\end{align}
where we have defined $T_{ab} =\frac{1 }{\kappa}\lrbrk{G_{ab} - \kappa T^{\mathrm{EM}}_{ab} }$ as the non-electromagnetic stress energy tensor, and $j^a = \nabla_b F^{ab}$ is the charge-current of the Maxwell sources.

We similarly obtain the Noether charges and constraints for all higher-derivative terms.  The results are presented below.

\paragraph{$\bm L_1$}
Variation of $\Lf{1}$ yields
\begin{equation}
\delta \Lf{1}  = \delta g_{ab} (E_1)^{ab} \vf 
+ \df{\bf \Theta}_1 \,,
\end{equation}
where we have defined
\begin{equation}
(E_1)^{ab}  = \frac{1}{2} g^{ab} R^2 -2R R^{ab} + 2 \nabla^b \nabla^a R - 2 g^{ab}\nabla_c \nabla^c R\,.
\end{equation}
The Noether charge associated with the vector field $\xi^a$ is
\begin{equation}
(Q^1_\xi)_{ab} = \epsilon_{abcd} \lrbrk{-4 \xi^c \, \nabla^d R + 2 R\,\nabla^d \xi^c} \,.
\end{equation}
The constraints are given by
\begin{equation}
C_{bcda} = -2 \epsilon_{ebcd} \, {\lrbrk{E_1}^{e}}_a \,.
\end{equation}

\bigskip

\paragraph{$\bm L_2$}
Variation of $\Lf{2}$ yields
\begin{equation}
\delta \Lf{2}  = \delta g_{ab} (E_2)^{ab} \vf 
+ \df {\bf \Theta}_2 \,,
\end{equation}
where we have defined
\begin{equation}
(E_2)^{ab}  = \frac{1}{2}g^{ab}R_{cd}R^{cd} + \nabla_c\nabla^b R^{ac} + \nabla_c\nabla^a R^{bc}
               -g^{ab}\nabla_d \nabla_c R^{cd} - \nabla^c \nabla_c R^{ab}
               -2 R^{ac}{R^{b}}_c\,.
\end{equation}
%
%
The Noether charge associated with the vector field $\xi^a$ is
\begin{equation}
(Q^2_\xi)_{ab} =\epsilon_{abcd} \, \lrbrk{ 
4 \xi^{[f}\, \nabla^{c]}{R_f}^d + {R_f}^d \nabla^f \xi^c + {R_f}^c \nabla^d \xi^f
} \,.
\end{equation}
The constraints are given by
\begin{equation}
C_{bcda} = -2 \epsilon_{ebcd} {\lrbrk{E_2}^{e}}_a \,.
\end{equation}

\bigskip

\paragraph{$\bm L_3$}
Variation of $\Lf{3}$ yields
\begin{equation}
\delta \Lf{3}  = \delta g_{ab} c_3 (E_3)^{ab} \vf 
+ \df {\bf \Theta}_3 \,,
\end{equation}
where we have defined
\begin{equation}
(E_3)^{ab}  = \frac{1}{2}g^{ab} R^2 + 2 g^{ab} R_{cd} R^{cd} + 2 R^{ab}R
- 8 R_{cd}R^{acbd} + 2 \nabla^b \nabla^a R - 4\Box R^{ab}   \,.
\end{equation}
%
%
The Noether charge associated with the vector field $\xi^a$ is
\begin{equation}
(Q^3_\xi)_{ab} = \epsilon_{abcd} \, \lrbrk{ -4 \xi^e \nabla_f {R_e}^{fcd} +2 {R_{ef}}^{cd} \nabla^f\xi^e}\,.
\end{equation}
The constraints are given by
\begin{equation}
C_{bcda} = -2 \epsilon_{ebcd} {\lrbrk{E_3}^{e}}_a \,.
\end{equation}

\bigskip

\paragraph{$\bm L_4$}

Variation of $\Lf{4}$ yields
\begin{equation}
\delta \Lf{4}  = \delta g_{ab} (E^{g}_4)^{ab} \vf  + \delta A_{a} (E^{A}_4)^{a} \vf
+ \df {\bf \Theta}_4 \,,
\end{equation}
where we have defined the equation of motions for $g_{ab}$ and $A_a$ respectively as 
\begin{align}
(E^{g}_4)^{ab} & = \lrsbrk{ - R^{ab} + \frac{1}{2} g^{ab} R -g^{ab} \nabla^2 + \nabla^{(a}\nabla^{b)} } F^2 - 2 R {F^{ac} {F_{b}}^{c}} \,, \\ 
(E^{A}_4)^{a} &= 4 \nabla_b \lrbrk{R F^{ab}} \,.
\end{align}
%
%
The Noether charge associated with the vector field $\xi^a$ is
\begin{equation}
(Q^4_\xi)_{ab} = \epsilon_{abcd} \lrbrk{F^2 \nabla^d \xi^c - 2\xi^c \nabla^d F^2 + 2 R F^{cd} A_e \xi^e}\,.
\end{equation}
The constraints are given by
\begin{equation}
C_{bcda} = -2 \epsilon_{ebcd} {(E^{g}_4)^{e}}_a - \epsilon_{ebcd} (E^{A}_4)^e A_a\,.
\end{equation}

\bigskip

\paragraph{$\bm L_5$}

Variation of $\Lf{5}$ yields
\begin{equation}
\delta \Lf{5}  = \delta g_{ab} (E^{g}_5)^{ab} \vf  + \delta A_{a} (E^{A}_5)^{a} \vf
+ \df {\bf \Theta}_5 \,,
\end{equation}
where we have defined the equation of motions for $g_{ab}$ and $A_a$ respectively as 
\begin{align}
(E^{g}_5)^{ab}  &= 
2F^{(bc} {F_c}^d {R^{a)}}_d - F^{ac}F^{bd} R_{cd} + \frac{1}{2} {F_c}^e F^{cd} g^{ab} R_{de}  \\ \nn
&- \nabla^{(a} F^{b)c}\nabla_d {F_c}^d - F^{cd} \nabla_d \nabla^{(a} {F^{b)}}_c - F^{(bc} \nabla_d \nabla^{a)} {F_c}^d - F^{(bc} \Box {F^{a)}}_c \\ \nn 
& -\nabla^{(b} F_{cd} \nabla^{d} F^{a)c} - F^{cd} g^{ab} \nabla_{(d} \nabla_{e)} {F_c}^e - \nabla_d {F^b}_c \nabla^d F^{ac} \\ \nn
&+\frac{1}{2} g^{ab} \nabla_c F^{cd} \nabla_e {F_d}^e - \frac{1}{2} g^{ab} \nabla_d F_{ce} \nabla^{e} F^{cd}
\,, \\
(E^{A}_5)^{a} & = 2 \nabla_c \lrbrk{ R^{bc} {F^a}_b + F^{bc} {R^a}_b
} \,.
\end{align}
The Noether charge associated with the vector field $\xi^a$ is
\begin{equation}
(Q^5_\xi)_{ab} = \epsilon_{abcd} \,\lrsbrk{
-2 \xi^e A_e F^{fc}{R_f}^d - 2 \xi^c F^{f(e} \nabla_e {F_f}^{d)}
+ \xi^e \nabla^d \lrbrk{F^{fc}F_{ef}} + {F_f}^d {F_e}^f \nabla^{[c}\xi^{e]}
} \,.
\end{equation}
The constraints are given by
\begin{equation}
C_{bcda} = -2 \epsilon_{ebcd} {(E^{g}_5)^{e}}_a - \epsilon_{ebcd} (E^{A}_5)^e A_a\,.
\end{equation}

\bigskip

\paragraph{$\bm L_6$}

Variation of $\Lf{6}$ yields
\begin{equation}
\delta \Lf{6}  = \delta g_{ab} (E^{g}_6)^{ab} \vf  + \delta A_{a} (E^{A}_6)^{a} \vf
+ \df {\bf \Theta}_6 \,,
\end{equation}
where we have defined the equation of motions for $g_{ab}$ and $A_a$ respectively as 
\begin{align}
(E^{g}_6)^{ab}  &= \frac{1}{2} F^{cd} F^{ef} g^{ab} R_{cdef} - 3 F^{(ac}F^{de}{R^{b)}}_{cde} \\ \nn
&- 2 F^{(ac}\nabla_c \nabla_d F^{b)d} - 2 F^{(ac}\nabla_d \nabla_c F^{b)d} - 4 \nabla_c F^{(ac} \nabla_d F^{b)d}
\,, \\ 
(E^{A}_6)^{a} &= 4 \nabla_d \lrbrk{ F^{bc} {R^{ad}}_{bc} }\,.
\end{align}
The Noether charge associated with the vector field $\xi^a$ is
\begin{equation}
(Q^6_\xi)_{ab} = \epsilon_{abcd} \, \lrsbrk{2 \xi^e A_e F^{fg}{R_{fg}}^{cd} - 2 \xi^e \nabla_f \lrbrk{F^{cd} {F_e}^f} + F^{cd} F_{ef} \nabla^f \xi^e
}
\,.
\end{equation}
The constraints are given by
\begin{equation}
C_{bcda} = -2 \epsilon_{ebcd} {(E^{g}_6)^{e}}_a - \epsilon_{ebcd} (E^{A}_6)^e A_a\,.
\end{equation}

\bigskip

\paragraph{$\bm L_7$}

Variation of $\Lf{7}$ yields
\begin{equation}
\delta \Lf{7}  = \delta g_{ab} (E^{g}_7)^{ab} \vf  + \delta A_{a} (E^{A}_7)^{a} \vf
+ \df {\bf \Theta}_7 \,,
\end{equation}
where we have defined the equation of motions for $g_{ab}$ and $A_a$ respectively as 
\begin{align}
(E^{g}_7)^{ab} & = \frac{1}{2} g^{ab}  F^2  F^2 - 4 F^{ac}  {F^b}_c  F^2\,, \\ 
(E^{A}_7)^{a} &=  8 \nabla_b \lrbrk{ F^{ab} F^2}\,.
\end{align}
The Noether charge associated with the vector field $\xi^a$ is
\begin{equation}
(Q^7_\xi)_{ab} = \epsilon_{abcd} \lrbrk{ 4 \xi^e A_e F^{cd} F^2}
\,.
\end{equation}
The constraints are given by
\begin{equation}
C_{bcda} = -2 \epsilon_{ebcd} {(E^{g}_7)^{e}}_a - \epsilon_{ebcd} (E^{A}_7)^e A_a\,.
\end{equation}

\bigskip

\paragraph{$\bm L_8$}

Variation of $\Lf{8}$ yields
\begin{equation}
\delta \Lf{8}  = \delta g_{ab} (E^{g}_8)^{ab} \vf  + \delta A_{a} (E^{A}_8)^{a} \vf
+ \df {\bf \Theta}_8 \,,
\end{equation}
where we have defined the equation of motions for $g_{ab}$ and $A_a$ respectively as 
\begin{align}
(E^{g}_8)^{ab}  &= \frac{1}{2} g^{ab} {F_c}^d {F_d}^e {F_e}^f {F_f}^c -4 F^{ac} F^{bd}  {F_c}^e F_{de}\,, \\ 
(E^{A}_8)^{a} & = - 8 \nabla_d \lrbrk{ {F^a}_b  {F^b}_c F^{cd}
} \,.
\end{align}
The Noether charge associated with the vector field $\xi^a$ is
\begin{equation}
(Q^8_\xi)_{ab} = \epsilon_{abcd} \lrbrk{ 4 \xi^e A_e {F_f}^d {F_g}^c F^{gf}
}\,.
\end{equation}
The constraints are given by
\begin{equation}
C_{bcda} = -2 \epsilon_{ebcd} {(E^{g}_8)^{e}}_a - \epsilon_{ebcd} (E^{A}_8)^e A_a\,.
\end{equation}

\bigskip

   Finally, the above results can be summarized in the following compact form:

\be
({\bf Q}_{\xi})_{c_3 c_4}=\epsilon_{ab c_3 c_4}\left( M^{abc}\,\xi_c- E^{abcd} \,\nabla_{[c}\;\xi_{d]}\right) \,, 
\ee
where
\be
M^{abc} \equiv -2 \nabla_d E^{abcd} + E_F^{ab}A^c\,,
\ee
and 
\be
({\bf C}^d)_{abc}=\epsilon_{eabc}(2E^{pqre}R_{pqr}^{\;\;\;\;d}+4\nabla_f\nabla_h E^{efdh}+2E^{eh}_F{F^d}_h-2A^d\nabla_hE_F^{eh} -g^{ed} \Lf{})
\ee
with
\be
E^{abcd}\equiv {\delta \Lf{} \over \delta R_{abcd}},\qquad E^{ab}_F \equiv {\delta \Lf{} \over \delta F_{ab}}\,.
\ee 
 
\section{Proof that constant area direction is along the extremality curve}\label{App-D}

  \begin{figure}
  \centering
  \includegraphics[width=0.8\linewidth]{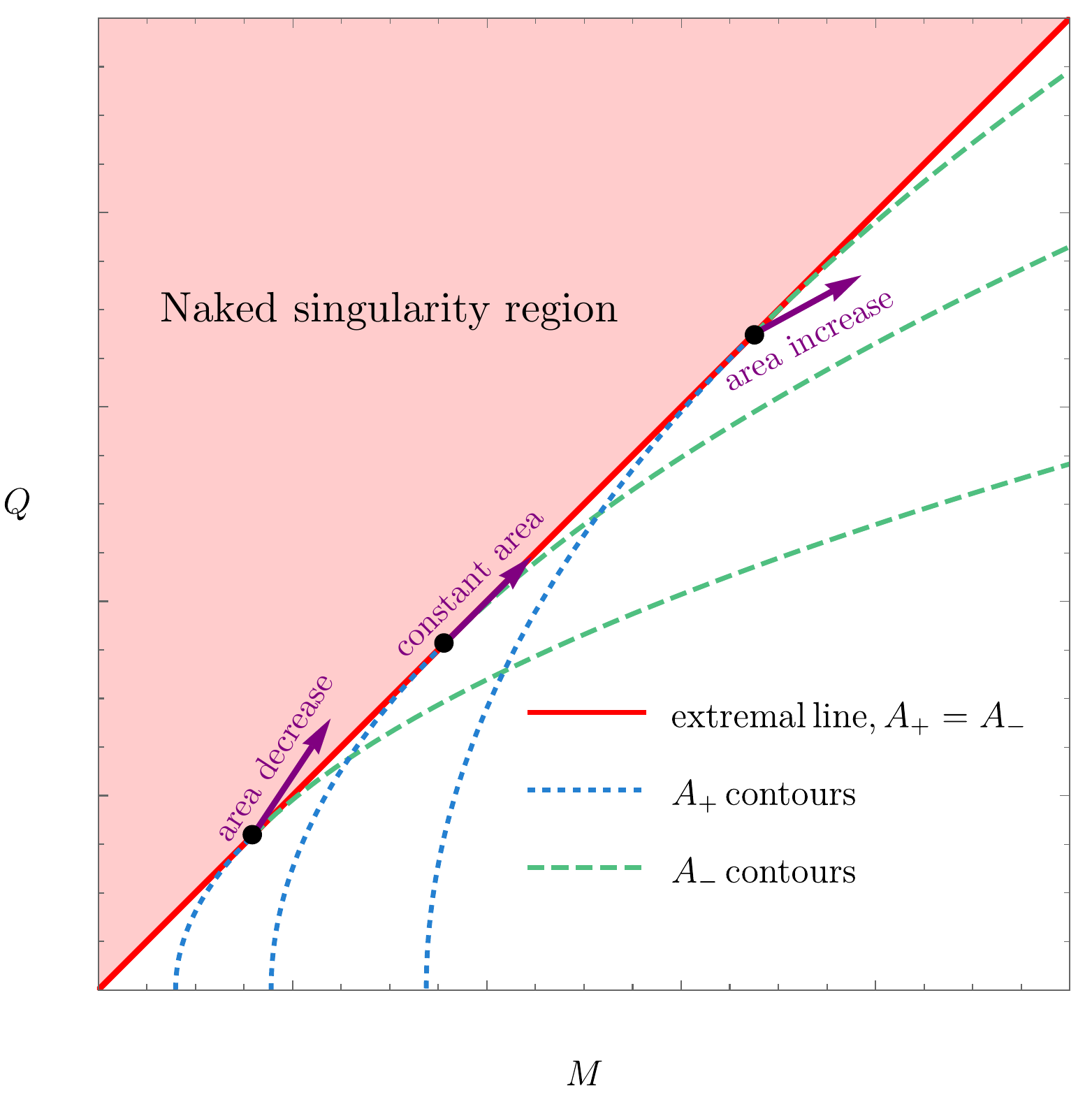}
  \caption{Extremality contour and constant area contours.  Extremal black holes live on the red solid line which divides the whole parameter space into the naked singularity region and the non-extremal black hole region.  The constant area contours are always tangent to the extremal line. A small perturbation around an extremal point then shifts the spacetime to one of the following: (i) a naked singularity when the horizon area is decreased; (ii) another extremal solution when the area is unchanged; and (iii) a nonextremal black hole when the area is increased. }
  \label{fig:contour}
  \end{figure}

  Suppose the radius, hence area $A$ of the horizon is determined implicitly by the following equation
  \begin{equation}
  \label{eq:area}
  F (M,Q,A) =0\,.
  \end{equation}
Extremality condition requires, in addition, that
\begin{equation}
\label{eq:ext}
\partial_A F(M,Q,A)=0\,.
\end{equation}
This is because the two roots of $1/g_{rr}$ coincide at this location. 

Extremal black holes is a one-parameter family, with $Q_{\rm ext}(M)$, $A_{\rm ext}(M)$ determined jointly by Eqs.~\eqref{eq:area} and \eqref{eq:ext}.   In practice, when $Q<Q_{\rm ext}(M)$, we will have contours of constant $A$ (as shown in Fig.~\ref{fig:contour}), determined by 
\begin{equation}
\partial _M F dM + \partial_Q F dQ = 0\,,
\end{equation}
or
\begin{equation}
\left(dQ/dM\right)_A = -\partial_M F/\partial_Q F\,.
\end{equation}

On the other hand, we can find out the direction of the extremality curve in the $(M,Q,A)$ space.  The tangent vector satisfies 
\begin{equation}
\partial_M F \Delta M + \partial_M F \Delta Q + \partial_A F \Delta A =0 \,.
\end{equation}
However, because we have $\partial_A F$ on that curve, we have $\partial_A F=0$ and  also 
\begin{equation}
\left(dQ/dM\right)_{\rm ext} = -\partial_M F/\partial_Q F\,.
\end{equation}
This means, on the extremality contour, the direction at which area remains constant is the same as the contour itself.  This does not mean that the contour all has the same area --- instead, constant area contours reach the extremality contour in a tangential way, as shown in the figure.

\end{document}